\documentclass[a4paper,11pt]{PoS}

\usepackage{amsmath}
\usepackage{subfigure}

\newcommand{\be}{\begin{equation}}
\newcommand{\ee}{\end{equation}}
\newcommand{\bea}{\begin{eqnarray}}
\newcommand{\eea}{\end{eqnarray}}

\title{Masses and decay constants of $D_{(s)}^*$ and $B_{(s)}^*$ mesons in Lattice QCD with $N_f = 2 + 1 + 1$ twisted-mass fermions}

\ShortTitle{Masses and decay constants of $D_{(s)}^*$ and $B_{(s)}^*$ mesons ...}

\author{Vittorio Lubicz\\
        Dip. di Matematica e Fisica, Universit\`a  Roma Tre \& INFN, Sezione di Roma Tre, Rome, Italy\\
        E-mail: \email{lubicz@fis.uniroma3.it}}

\author{Aurora Melis\\
        Dipartimento di Matematica e Fisica, Universit\`a  Roma Tre, Rome, Italy\\
        E-mail: \email{me.aurora.16@gmail.com}}

\author{Silvano Simula\\
        Istituto Nazionale di Fisica Nucleare, Sezione di Roma Tre, Rome, Italy\\
        E-mail: \email{simula@roma3.infn.it}\\
        \\
        \textbf{for the ETM Collaboration}}

\abstract{We present a lattice calculation of the decay constants and masses of $D_{(s)}^*$ and $B_{(s)}^*$ mesons using the gauge configurations produced by the European Twisted Mass Collaboration (ETMC)  with $N_f = 2 + 1 + 1$ dynamical quarks and at three values of the lattice spacing $a \sim 0.06 - 0.09$ fm. Pion masses are simulated in the range $m_{\pi} \sim 210 - 450$ MeV, while the strange and charm quark masses are close to their physical values. We computed the ratios of vector to pseudoscalar decay constants or masses for various values of the heavy-quark mass $m_h$ in the range $0.7 m_c^{phys} \lesssim m_h \lesssim 3 m_c^{phys}$. In order to reach the physical b-quark mass, we exploited the HQET prediction that, in the static limit of infinite heavy-quark mass, all the considered ratios are equal to one. We obtain: $ f_{D^*}/f_{D} = 1.078(36),$  $m_{D^*}/m_{D} = 1.0769(79)$, $f_{D^*_{s}}/f_{D_{s}} = 1.087(20)$, $m_{D^*_{s}}m_{D_{s}} = 1.0751(56)$, $f_{B^*}/f_{B} = 0.958(22)$, $m_{B^*}/m_{B} = 1.0078(15)$, $f_{B^*_{s}}/f_{B_{s}} = 0.974(10)$ and $m_{B^*_{s}}/m_{B_{s}} = 1.0083(10)$. Combining them with the corresponding experimental masses from the PDG and the pseudoscalar decay constants calculated by ETMC, we get: $f_{D^*} = 223.5(8.4)~\mathrm{MeV}$, $m_{D^*} = 2013(14)~\mathrm{MeV}$, $f_{D^*_{s}} = 268.8(6.6)~\mathrm{MeV}$, $ m_{D^*_{s}} = 2116(11)~\mathrm{MeV}$, $f_{B^*} = 185.9(7.2)~\mathrm{MeV}$, $m_{B^*} = 5320.5(7.6)~\mathrm{MeV}$, $f_{B^*_{s}} = 223.1(5.4)~\mathrm{MeV}$ and $m_{B^*_{s}}= 5411.36(5.3)~\mathrm{MeV}$.}

\FullConference{34th annual International Symposium on Lattice Field Theory\\
                 24-30 July 2016\\
                 University of Southampton, UK}

\begin{document}

\section{Introduction}
\label{intro}

The decay constants of $D_{(s)}^*$ and $B_{(s)}^*$ mesons are important ingredients in the phenomenological description of various processes, like semileptonic and non-leptonic decays of heavy hadrons. 
It is well known that in the limit of infinite heavy-quark mass the Heavy Quark Effective Theory (HQET) predicts that the ratios of vector (V) to pseudoscalar (PS) decay constants or masses are equal to one, i.e.~$\lim_{m_h \rightarrow \infty} (M_{H^*} / M_H) = 1$ and $\lim_{m_h \rightarrow \infty} (f_{H^*} / f_H) = 1$. 
When the heavy quark is either the charm or the beauty, the spin-flavor symmetry is broken and the above ratios deviate from one because of power corrections in $1 / m_h$.
Till now there are only few lattice calculations of the $D_{(s)}^*$ and $B_{(s)}^*$ decay constants using gauge configurations with $N_f = 2$ \cite{incriminato,sanfilippo_noi} and $N_f = 2 + 1 (+1)$ \cite{DavisD,DavisB} dynamical quarks.
These results exhibit a surprisingly non-negligible dependence on $N_f$.

In this contribution we present the results obtained for the V to PS ratios of decay constants and masses in the charm and beauty sectors using the  gauge ensembles generated by the European Twisted Mass Collaboration (ETMC) with $N_f = 2 + 1 + 1$ dynamical quarks \cite{Baron:2010bv,Baron:2010th,Baron:2011sf}. 
In the ETMC set-up the gluon interactions are described by the Iwasaki action, while the fermions are regularised in the maximally twisted-mass (Mtm) Wilson lattice formulation. 
We considered three values of the lattice spacing, namely $a = 0.0885(36), 0.0815(30)$ and $0.0619(18)$ fm, with the lowest simulated pion mass being equal to $\simeq 210$ MeV. 
The valence quark masses are chosen to be in the ranges: $3 m_{ud}^{phys} \lesssim m_{ud} \lesssim 12 m_{ud}^{phys}$, $0.7 m_s^{phys} \lesssim m_s \lesssim 1.2 m_s^{phys}$ and $0.7 m_c^{phys} \lesssim m_c \lesssim 1.1 m_c^{phys}$. 
To extrapolate up to the b-quark sector we have also considered higher values of the valence heavy-quark mass in the range $1.1 m_c^{phys} \lesssim m_h \lesssim 3 m_c^{phys} \approx 0.7 m_b^{phys}$. 
The lattice scale was determined using the experimental value of $f_{\pi^+}$ \cite{Carrasco:2014cwa}, while the physical up/down, strange, charm and bottom quark masses were obtained \cite{Carrasco:2014cwa,Bussone:2016iua} by using the experimental values for $m_\pi$, $m_K$, $m_D$ and $m_B$, respectively.

In Ref.~\cite{Carrasco:2014cwa} eight branches of the analysis were adopted. 
They differ in: 
~ i) the continuum extrapolation adopting for the scale parameter either the Sommer parameter $r_0$ or the mass of a fictitious PS meson made up of strange(charm)-like quarks; 
~ ii) the chiral extrapolation performed with fitting functions chosen to be either a polynomial expansion or a Chiral Perturbation Theory (ChPT) ansatz in the light-quark mass;
~ iii) the choice between the methods M1 and M2, which differ by $O(a^2)$ effects, used to determine in the RI'-MOM scheme the mass renormalization constant (RC) $Z_m = 1 / Z_P$. 
In the present analysis we made use of the input parameters corresponding to each of the eight branches of Ref.~\cite{Carrasco:2014cwa}.

\section{Extraction of masses and decay constants}
\label{analysis}

The decay constants of V and PS mesons are defined in terms of the matrix elements
 \bea
    \label{fV}
    \langle 0 | \overline{h} \gamma_\mu \ell | H^*_\ell(\vec{p}, \lambda) \rangle & = & f_{H^*_{\ell}} m_{H^*_{\ell}} \epsilon_\mu^\lambda ~ , \\[2mm]
    \label{fP}
    \langle 0 | \overline{h} \gamma_5 \ell | H_\ell(\vec{p}) \rangle & = & \langle 0 | \partial_\mu \left(\overline{h} \gamma_\mu \gamma_5 \ell \right) 
                                                                                            | H_\ell(\vec{p} \rangle / (m_h + m_\ell) = f_{H_{\ell}} m_{H_{\ell}}^2 / (m_h + m_\ell) ~ ,
 \eea
where $m_h$ and $m_\ell$ are the heavy- and light-quark masses with $h = \{c, b\}$ and $\ell = \{ud, s\}$, and $\epsilon_\mu^\lambda$ is the vector meson polarization. 
Ground-state masses and decay constants can be determined by studying two-point correlation functions at large time distances, viz.
 \be
    \label{CVt}
     C_V(t) = \frac{1}{3} \langle \sum_{i,\vec{x}} V_i(\vec{x}, t) V_i^{\dagger}(0,0) \rangle \xrightarrow[t \geq t_{\mathrm{min}}]{}
                          \sum_i | \langle 0| V_i(0) |H^*_\ell(\lambda) \rangle |^2 ~ \frac{\cosh[m_{H^*_\ell}(T/2 - t)]}{3m_{H^*_\ell}} 
                          e^{-m_{H^*_\ell}T/2} ~ , 
 \ee
 \be
    \label{CPt}
    C_P(t) = \langle \sum_{\vec{x}} P(\vec{x}, t) P^{\dagger}(0,0) \rangle \xrightarrow[t \geq t_{\mathrm{min}}]{}
                   | \langle 0| P(0) |H_\ell \rangle |^2 ~ \frac{\cosh[m_{H_\ell}(T/2 - t)]}{m_{H_\ell}} e^{-m_{H_\ell}T/2} ~ , 
 \ee
where $t_{\mathrm{min}}$ stands for the minimum time at which the ground state can be considered well isolated.
In Eq.~(\ref{CVt}) $V_i \equiv Z_A \overline{h} \gamma_i \ell$ is the local vector current, which in our Mtm setup renormalizes multiplicatively with the RC $Z_A$, while in Eq.~(\ref{CPt}) $P \equiv  (m_h + m_\ell) Z_P \overline{h} \gamma_5 \ell = (\mu_h + \mu_\ell) \overline{h} \gamma_5 \ell$, where  $\mu_h$ and $\mu_\ell$ are bare quark masses, is the pseudoscalar interpolating field, which  in our Mtm setup is renormalization group invariant and does not require any RC. 
The meson masses $m_{H_\ell}$ and $m_{H^*_\ell}$ are extracted from the plateaux of the effective mass at large $t \geq t_{\mathrm{min}}$, while the correlation functions (\ref{CVt}) and (\ref{CPt}) for $t \geq t_{\mathrm{min}}$ contain the required matrix elements. 

In Eqs.~(\ref{CVt}-\ref{CPt}) we considered local source and sink operators, but we analyzed also the whole set of four correlation functions given by the combinations of local interpolating operators with those obtained from a Gaussian smearing procedure in both the sink and the source, namely $C^{LL}_{P,V}, C^{LS}_{P,V}, C^{SL}_{P,V}$ and $C^{SS}_{P,V}$, where $L$ and $S$ denote local and smeared operators, respectively. 
It is straightforward to check that the required local matrix elements in Eqs.~(\ref{fV}-\ref{fP}) can be extracted from the $LL$ correlation functions as well as from an appropriate combination of the $SL$ and $SS$ ones, that is $C^{SL}_{P,V}(t)/\sqrt{C^{SS}_{P,V}(t)}$.

\begin{figure}[htb!]
\centering{
\scalebox{1}{\includegraphics[width=7.5cm]{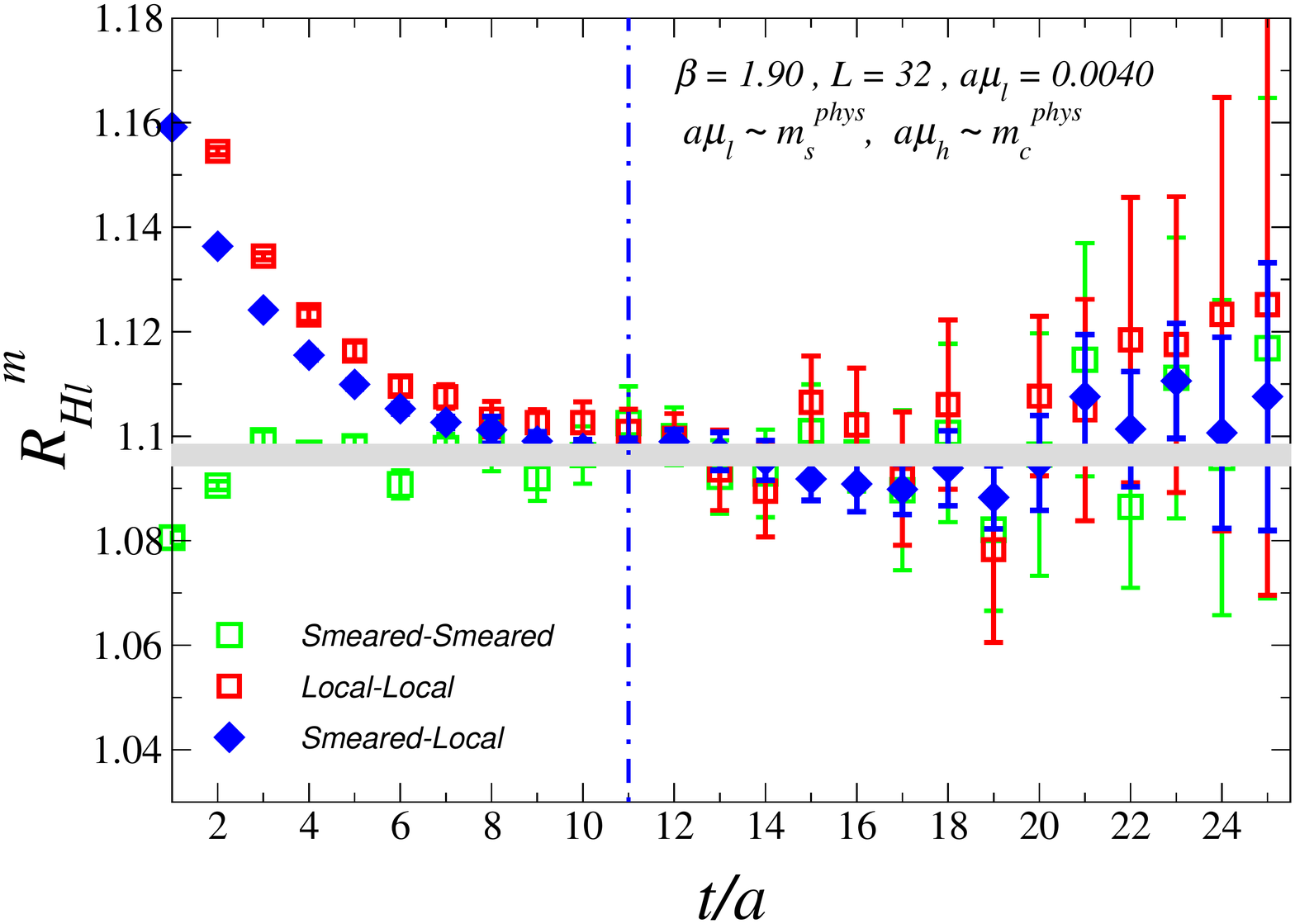}}
\scalebox{1}{\includegraphics[width=7.5cm]{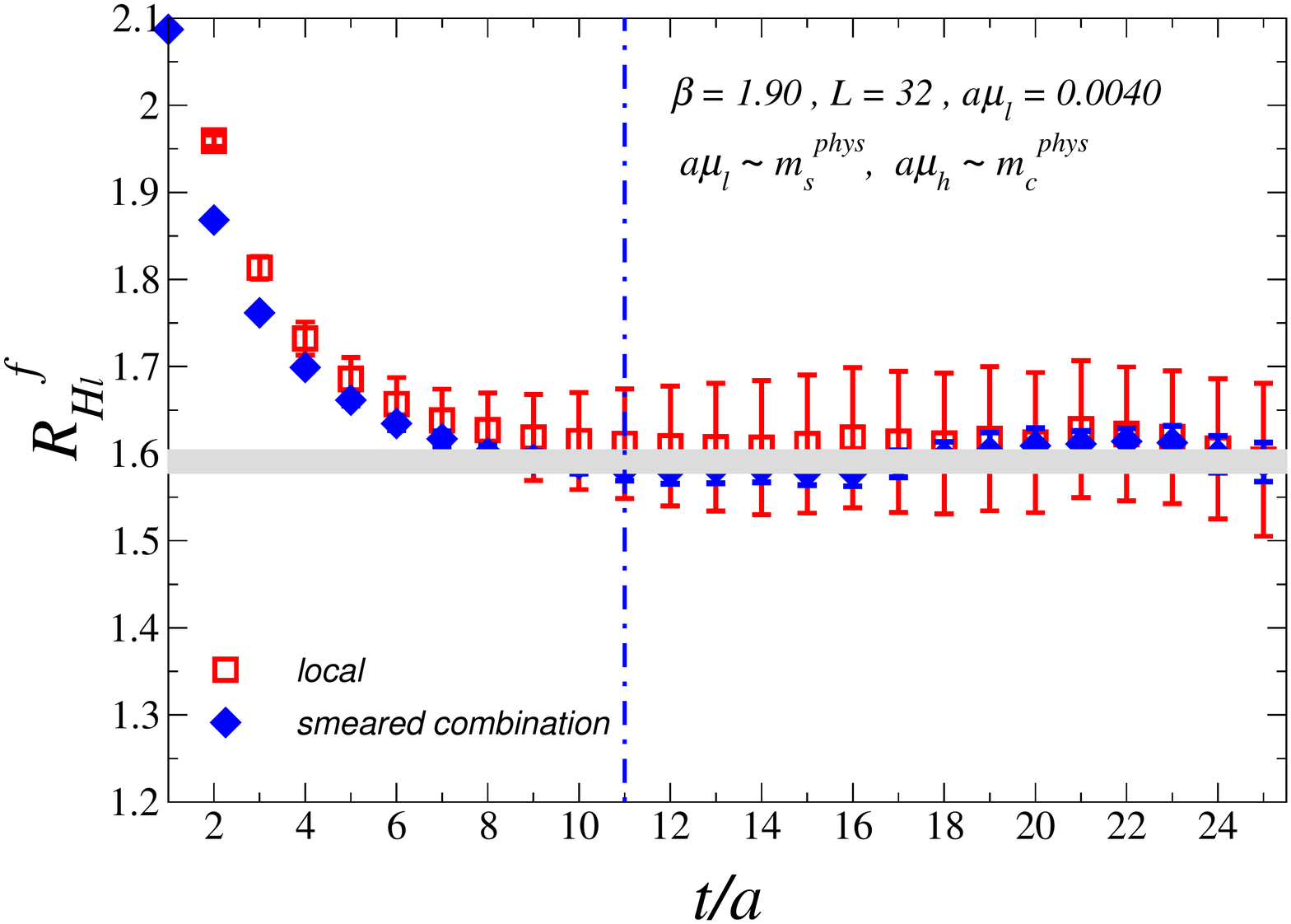}}
}
\vspace{-0.75cm}
\caption{\small \it The ratios $R^m_{H_\ell}$ and $R^f_{H_\ell}$ (see Eq.~(\protect\ref{FeM})) using only local (red points) or also Gaussian smeared operators (blue points). The value of $t_{min}$ is shown as the vertical dot-dashed line. The grey bands are the mass and the decay constant, obtained as constant fits in the plateaux regions.}
\label{fig:LLvsSL}
\end{figure}

For the reasons explained in the Introduction we have considered the following ratios 
 \be
    R_{H_\ell}^m = m_{H_\ell^*}/m_{H_\ell} \hspace{5mm} \mathrm{and} \hspace{5mm} R_{H_\ell}^f = f_{H_\ell^*}/f_{H_\ell} ~ .
    \label{FeM}
 \ee 
In Fig.~\ref{fig:LLvsSL} we show an example of the above quantities by comparing the extraction from Gaussian-smeared and/or local correlation functions. 
The smearing techniques allows plateaux to start at earlier time distances, and the $SL$ correlation functions exhibit the best signal to noise ratio. The value of $t_{min}$ corresponds to the smallest time distance where the effective masses obtained from $SL$ and $SS$ correlation functions intercept each other.

\subsection{The $D_{(s)}^*$-meson masses and decay constants}
\label{sec:FMD}

\begin{figure}[htb!]
\centering{
\scalebox{1}{\includegraphics[width=7.5cm]{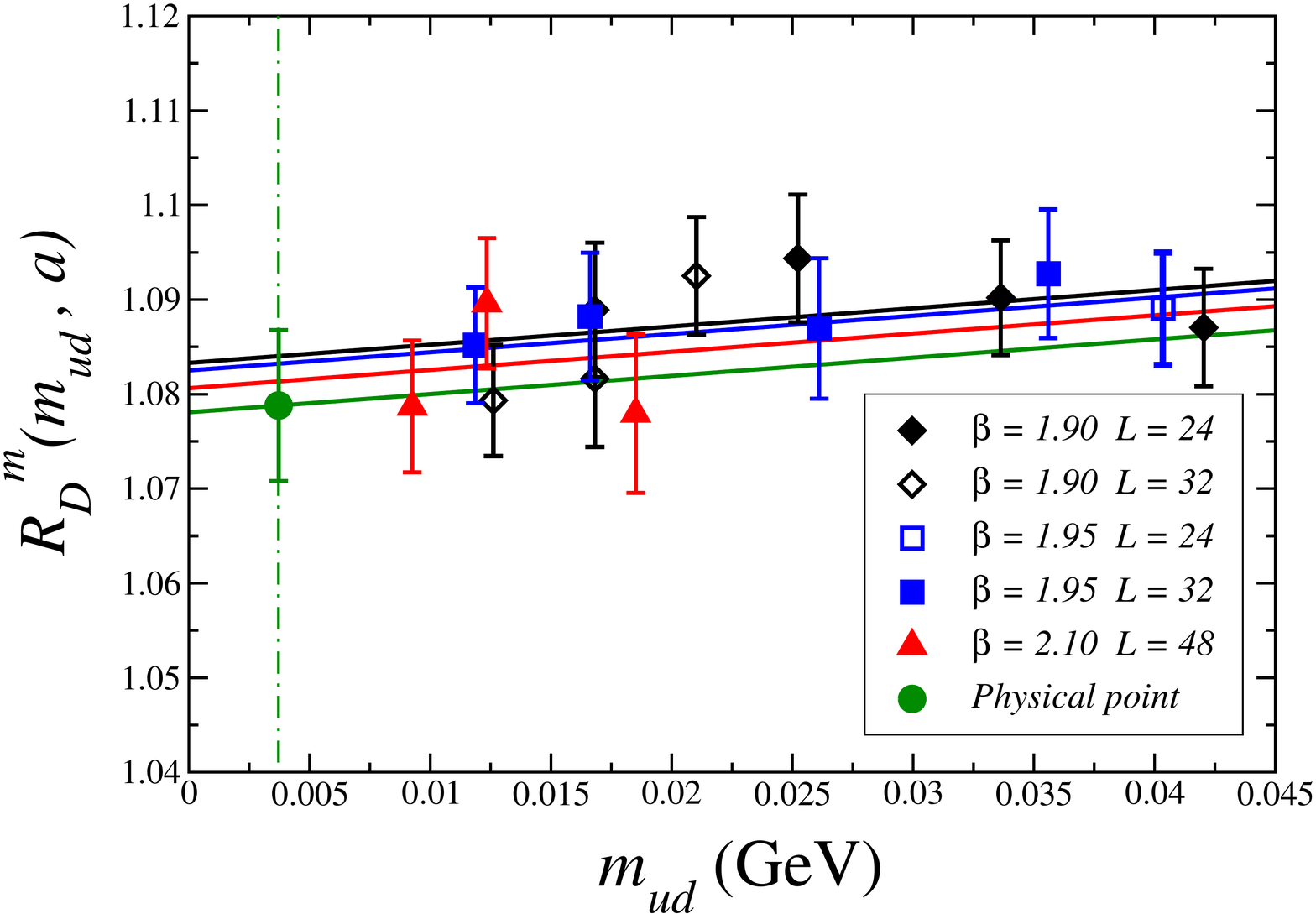}}
\scalebox{1}{\includegraphics[width=7.5cm]{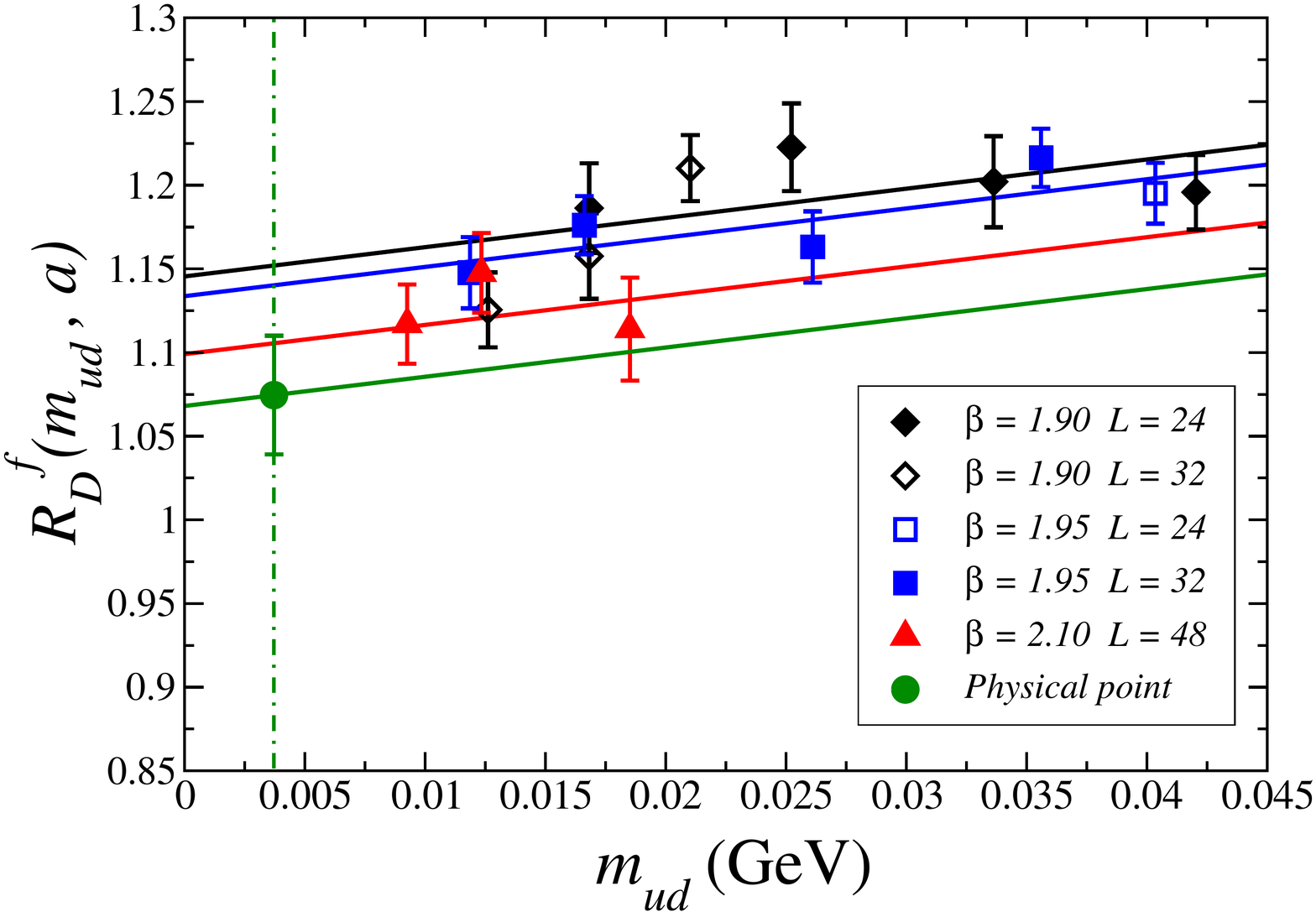}}
}
\vspace{-0.75cm}
\caption{\small  \it Chiral and continuum extrapolations of $R_D^m$ and $R_D^f$ based on the polynomial fit (\protect\ref{fit}) with $P_3 = P_4 = 0$. The green points represent the values at the physical point $m_{ud}^{phys}(\overline{MS},~2~\mbox{GeV}) = 3.70 (17)$ MeV \protect\cite{Carrasco:2014cwa}. Similar results hold as well in the case of the ratios $R_{D_s}^m$ and $R_{D_s}^f$.}
\label{fig:ContD} 
\end{figure}

We perform a smooth interpolation of the lattice data for the ratios $R_{H_\ell}^m$ and $R_{H_\ell}^f$ to the values of the physical strange and charm quark masses $m_s^{phys}(\overline{MS},~2~\mbox{GeV}) = 99.6(4.3)$ MeV and $m_c^{phys}(\overline{MS},~2~\mbox{GeV}) = 1.176(39)$ GeV~\cite{Carrasco:2014cwa}. 
The dependence of $R_{D_{(s)}}^m$ and $R_{D_{(s)}}^f$ on the renormalized up/down quark mass $m_{ud} = a\mu_{ud} / (aZ_P )$ and the lattice spacing $a$ is investigated by performing a combined chiral and continuum extrapolation, based on a polynomial expansion of the form
 \be
    R_{D_{(s)}}^{fit}(m_{ud}, a)= P_0 + P_1 m_{ud} + P_2 a^2 + P_3 m_{ud}^2 + P_4 a^4,
    \label{fit}
 \ee
where we have taken into account that for our Mtm setup the automatic $O(a)$-improvement implies that discretization effects involve only even powers of the lattice spacing. 
The results obtained with quadratic $m_{ud}^2$ and quartic $a^4$ terms have not been included in the final average (which therefore corresponds to $P_3  = P_4 = 0$), but they have been considered to estimate the uncertainty related to the chiral and continuum extrapolation, respectively. 
The latter ones are shown in Fig.~\ref{fig:ContD}, where the physical point corresponds to $m_{ud}^{phys}(\overline{MS},~2~\mbox{GeV}) = 3.70 (17)$ MeV~\cite{Carrasco:2014cwa}.

In this way at the physical point we get
 \bea
    m_{D^*}/m_D & = & 1.0769\,(71)_{stat} (30)_{input}(13)_{tmin} (8)_{disc} (5)_{chir}\,[79] ~ , \\
    m_{D^*_s}/m_{D_s} & = & 1.0751(49)_{stat} (27)_{input} (8)_{disc} (4)_{tmin} (2)_{chir}\,[56] ~ , \\
    f_{D^*}/f_{D} & = & 1.078\,(31)_{stat} (9)_{chir} (8)_{disc} (6)_{tmin} (5)_{input}\,[36] ~ , \\
    f_{D^*_s}/f_{D_s} & = & 1.087\,(16)_{stat} (7)_{disc} (6)_{input} (6)_{tmin} (5)_{chir}\,[20] ~ ,
 \eea
where total uncertainty (in the square brakets) is the sum in quadrature of the statistical and all the systematic uncertainties, which have been written in order of relevance for each of the ratios.
The various systematic uncertainties are estimated in the following way: 
~ i) the uncertainty labelled $tmin$ is computed by repeating the analysis with a value of $t_{\mathrm{min}}$ shifted by two units and taking the half difference with the final estimates;                                                            
~ ii) the chiral and discretization uncertainties, labelled respectively as $chir$ and $disc$, are obtained by considering either $P_3 \neq 0$ or $P_4 \neq 0$ in Eq.~(\ref{fit}) and taking again half of the difference with the final results; 
~ iii) the uncertainty labelled $input$ is given by the spread of the results over the eight branches of the input parameters of Ref.~\cite{Carrasco:2014cwa}.

By combining $R^m_{D_{(s)}}$ with the experimental values of the $D_{(s)}$-meson masses~\cite{PDG} we obtain 
\be
     m_{D^*} = 2013 ~ (14) ~ \mbox{MeV} \hspace{5mm} \mathrm{and} \hspace{5mm} m_{D_s^*} = 2116 ~ (11) ~ \mbox{MeV} ~ ,
\ee
that compare well with the experimental meson masses $m_{D^*}^{exp} = 2010.27 (5)$ MeV and $m_{D_s^*}^{exp} = 2112.1 (4)$ MeV \cite{PDG}. 

As for the decay constants, existing lattice calculations for $R^f_{D_{(s)}}$ have been carried out only with $N_f = 2 + 1$ and $N_f = 2$ dynamical quarks. 
The $N_f = 2 + 1$ estimate $f_{D_s^*} / f_{D_s} = 1.10 (2)$ \cite{DavisD} is in good agreement with our result, while the $N_f = 2$                                                                                                                                                                                                                                                                                                                                                                                                                                                                                                                                                                                                                                                                                                 results $f_{D^*} / f_D = 1.208 (27)$~\cite{sanfilippo_noi} and $f_{D_s^*} /f _{D_s} = 1.26( 3)$\cite{incriminato} are $\simeq 10\%$ larger than our predictions. 

Using the pseudoscalar decay constants calculated by ETMC in Ref.~\cite{leptonic} we get 
\be
     f_{D^*} = 223.5 ~ (8.7)~ \mbox{MeV} \hspace{5mm} \mathrm{and} \hspace{5mm} f_{D_s^*}= 268.8 ~ (6.5) ~ \mbox{MeV} ~ .
 \ee

\subsection{The $B_{(s)}^*$-meson masses and decay constants}
\label{sec:FfD}

\begin{figure}[htb!]
\centering{
\scalebox{1}{\includegraphics[width=7.5cm]{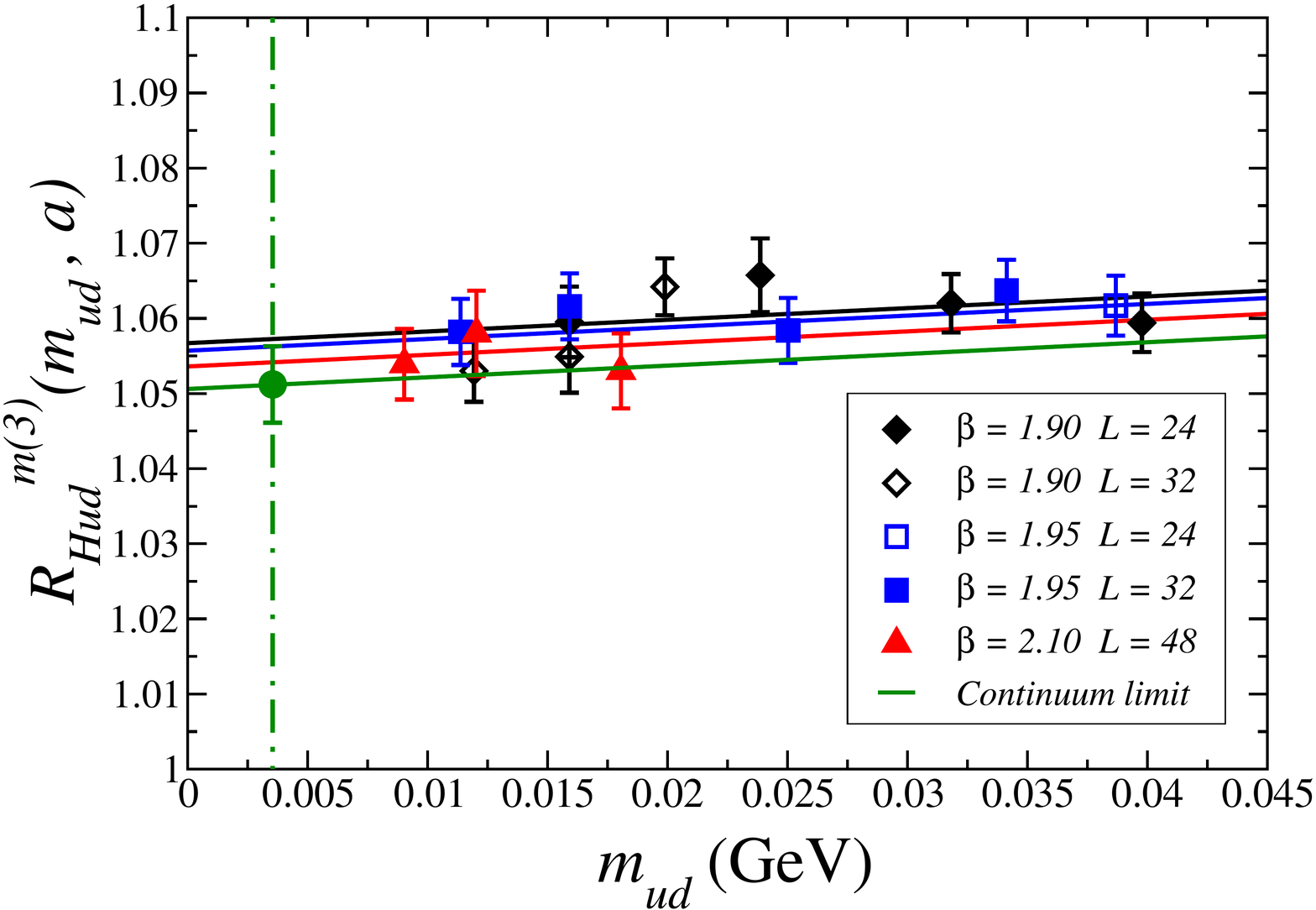}}
\scalebox{1}{\includegraphics[width=7.5cm]{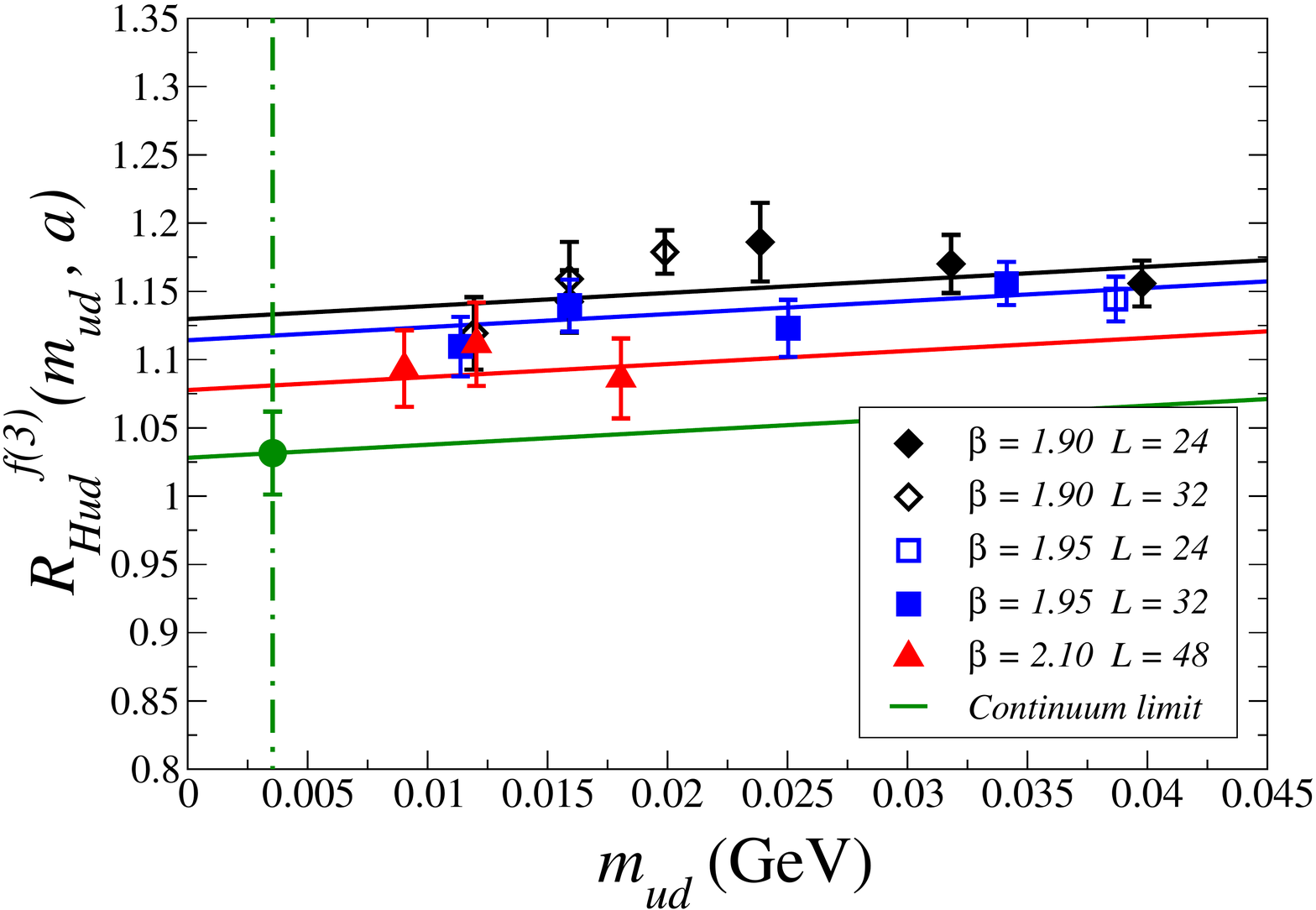}}
}
\vspace{-0.75cm}
\caption{\it \small Chiral and continuum extrapolations of $R^{m(k)}_{H_{ud}}$ and $R^{f(k)}_{H_{ud}}$ for $k = 3$, based on the polynomial fit (\protect\ref{fit}) with $P_3 = P_4 = 0$. Similar results hold as well in the case of $R^{m(k)}_{H_s}$ and $R^{f(k)}_{H_s}$.}
\label{fig:F_h} 
\end{figure}

We have computed the ratios $R^{m(k)}_{H_\ell}$ and $R^{f(k)}_{H_{\ell}}$ for a series of masses $\{m_h^{(k)}\} \geq m_c$ with $k = 1, ..., 8$. 
The results are extrapolated to the chiral and continuum limits, as shown in Fig.~\ref{fig:F_h}.

\begin{figure}[htb!]
\centering{
\scalebox{0.95}{\includegraphics[width=7.5cm]{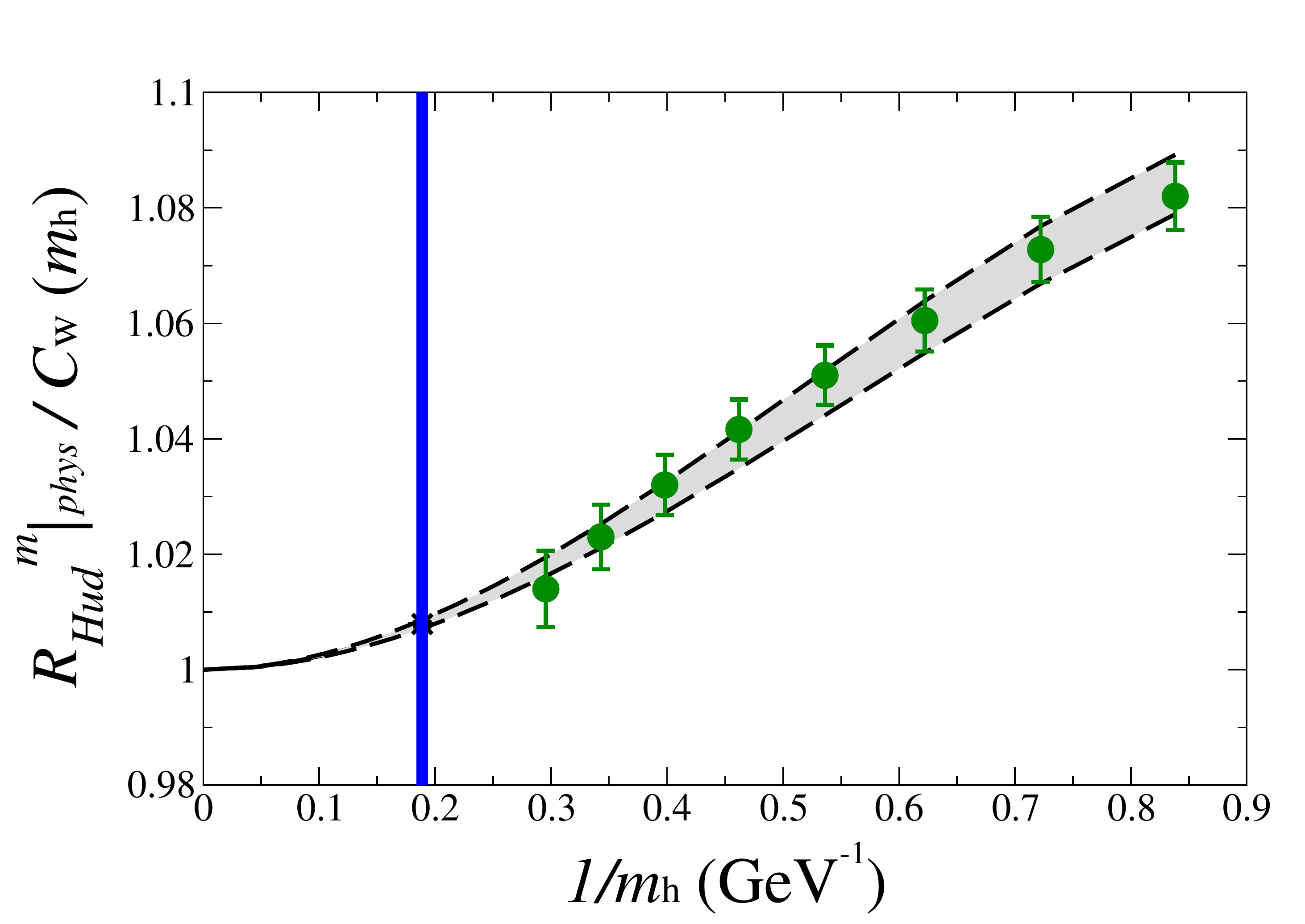}}
\scalebox{0.95}{\includegraphics[width=7.5cm]{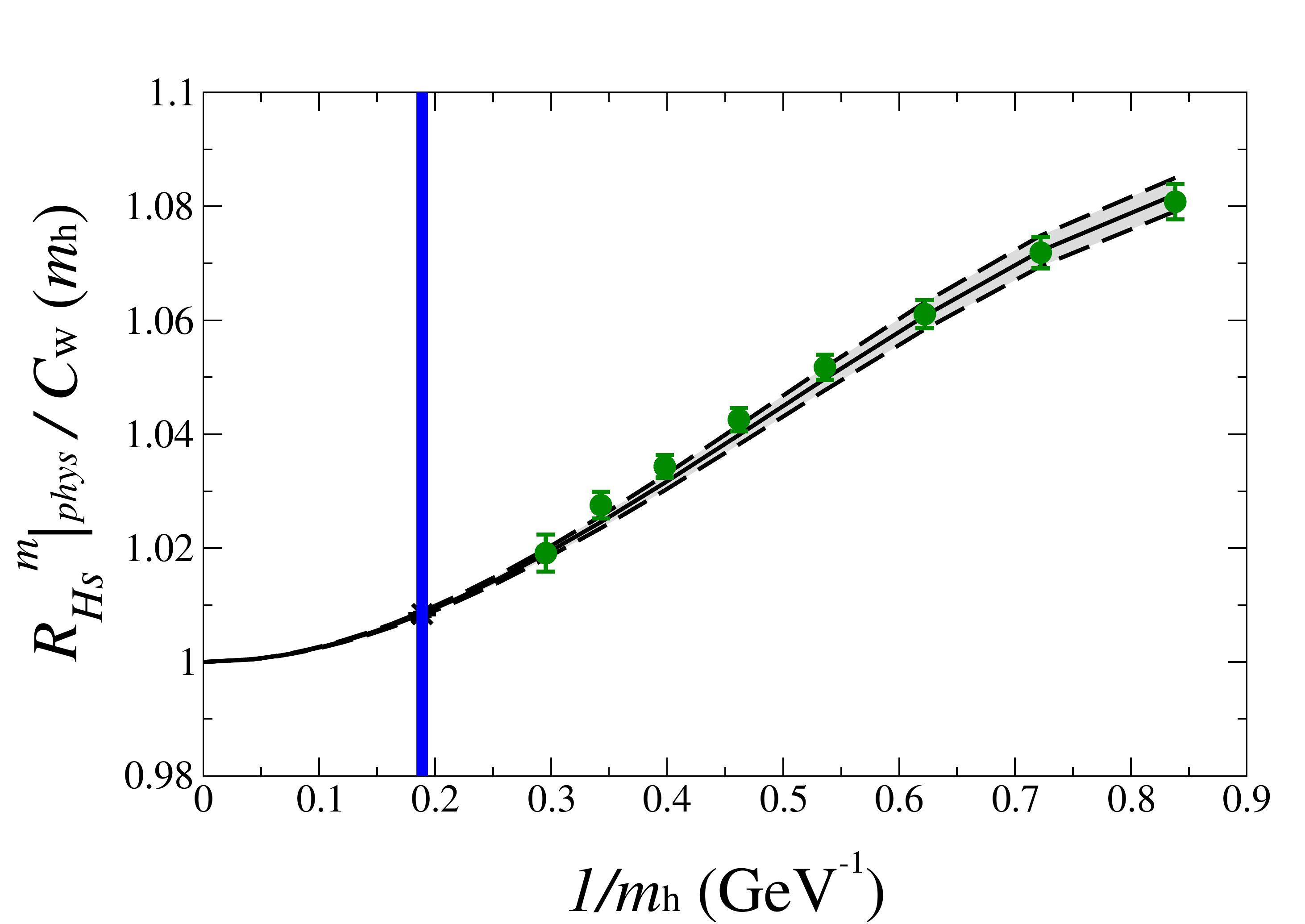}}
\scalebox{0.95}{\includegraphics[width=7.5cm]{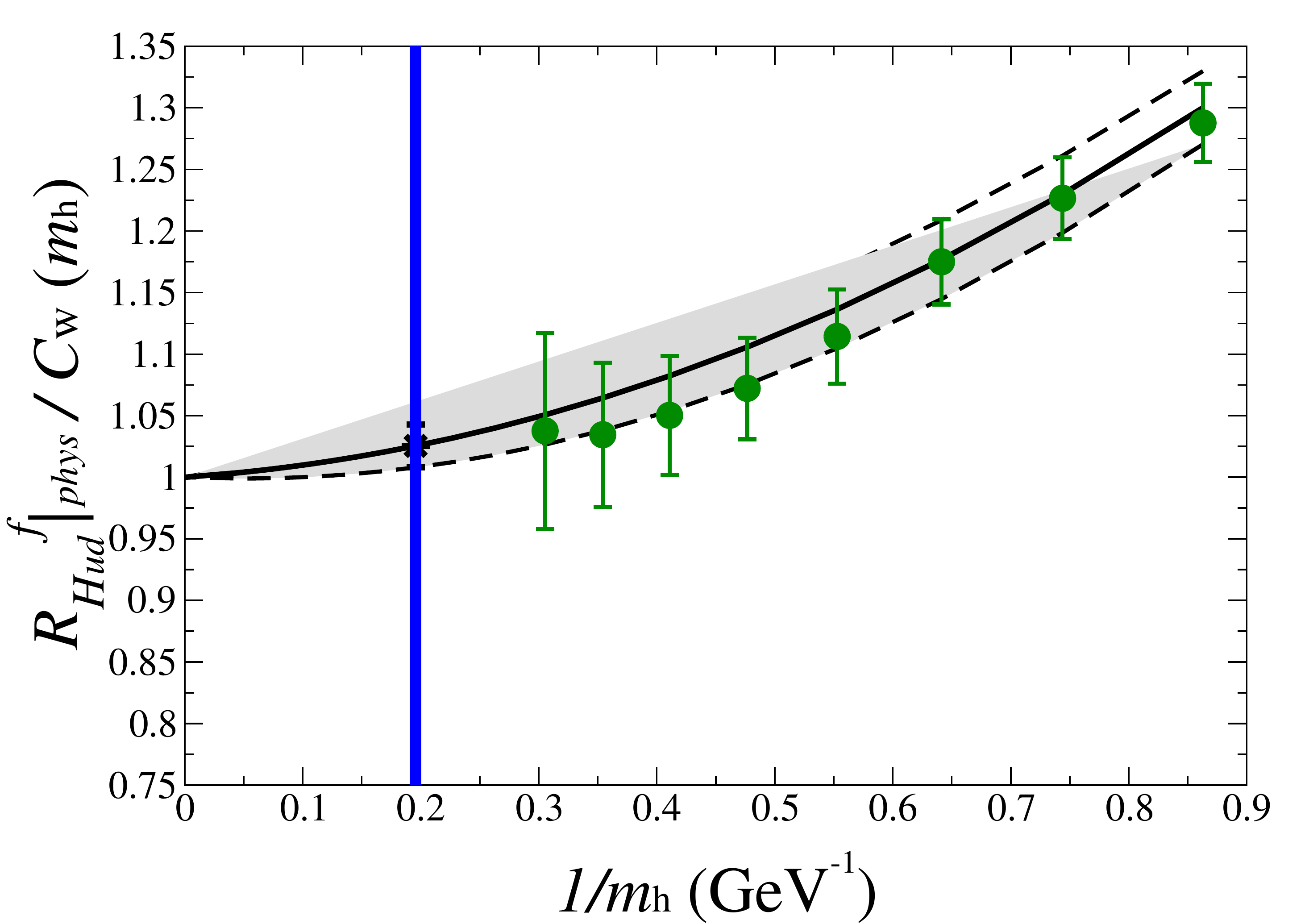}}
\scalebox{0.95}{\includegraphics[width=7.5cm]{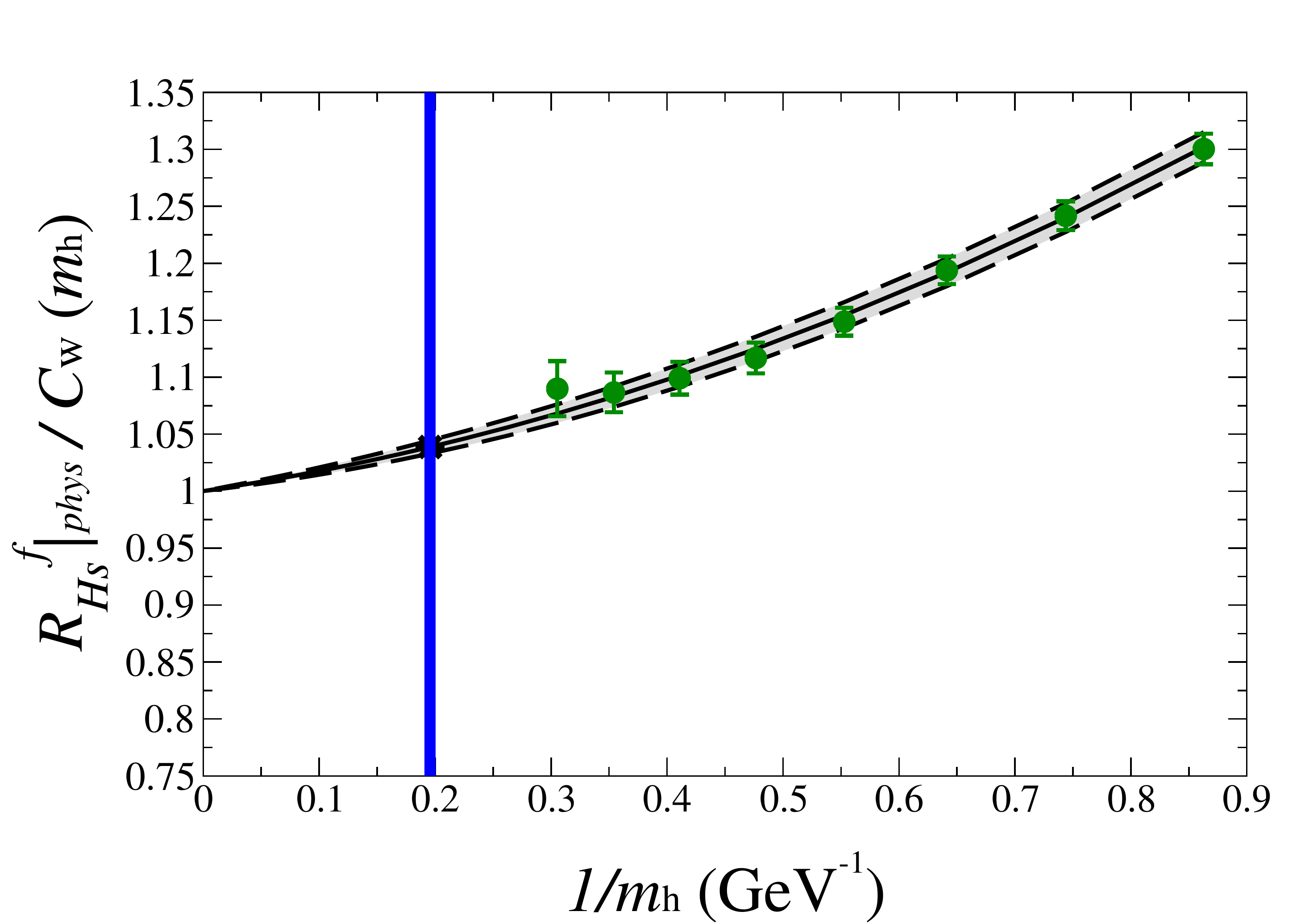}}
}
\vspace{-0.25cm}
\caption{\small  \it The dependence of $R_{H_{ud(s)}}^m|_{phys}$ and $\overline{R}_{H_{ud(s)}}^f|_{phys}$, extrapolated to the chiral and continuum limits, on the inverse heavy-quark mass $1 / m_h(\overline{MS},~2~\mbox{GeV})$. The fits are based on Eqs.~(\protect\ref{fith}) and the correlation matrix among the data points is accounted for. The vertical bands correspond to $1 / m_b^{phys}$ from Ref.~\protect\cite{Bussone:2016iua}.}
\label{fig:M_B}
\end{figure}

The HQET predicts that the ratios $R^m_{H_\ell}$ and $\overline{R}^f_{H_{\ell}} \equiv R^f_{H_\ell} / C_W(m_h)$, where $C_W(m_h)$ is the perturbative matching correction between full QCD and HQET (computed in Ref.~\cite{matching} up to next-to-next-leading order), are equal to one in the static heavy-quark limit, viz.
 \be
     \lim_{m_h \rightarrow \infty} R^m_{H_\ell} = 1 \hspace{5mm} \mathrm{and} \hspace{5mm} 
     \lim_{m_h \rightarrow \infty} \overline{R}^f_{H_{\ell}} = \lim_{m_h \rightarrow \infty} R^f_{H_\ell} / C_W(m_h) = 1 ~ .
     \label{constraint}
 \ee
Thus, we perform correlated polynomial fits in $1 / m_h$ imposing the static limit constraint, namely
 \bea
    R_{H_\ell}^m|_{\mathrm{phys}}^{fit} & = & 1 + \overline{D}_2 / m_h^2 + \overline{D}_3 / m_h^3  + \overline{D}_4 / m_h^4 ~ , \nonumber \\
    \overline{R}_{H_\ell}^f|_{\mathrm{phys}}^{fit} & = & 1 + D_1 / m_h + D_2 / m_h^2 + D_3 / m_h^3 ~ , 
     \label{fith} 
 \eea
where we have taken into account that, according to HQET, the linear term is absent in the case of the mass ratio (i.e., $\overline{D}_1 = 0$).
In Fig.~\ref{fig:M_B} the interpolations of the various ratios in the inverse heavy-quark mass are shown together with the results at the $b$-quark physical point obtained using the value $m_b^{phys}$ from Ref.~\cite{Bussone:2016iua}.

Our final results for $B_{(s)}^*$ mesons are
 \bea
    \label{ris_B}
    m_{B^*}/m_B & = &1.0078\,(8)_{stat}(8)_{chir} (7)_{tmin}(5)_{disc}(2)_{input}\,[14] ,\\
    m_{B^*_s}/m_{B_s} & = & 1.0083\,(6)_{stat}(7)_{chir}(6)_{disc}(3)_{tmin}(2)_{input}\,[11] , \\
    f_{B^*}/f_B & = & 0.958\,(18)_{stat} (10)_{disc}(6)_{chir}(5)_{tmin}(2)_{input}\,[22] , \\
    f_{B^*_{s}}/f_{B_{s}} & = & 0.974\,(7)_{stat}(6)_{disc}(3)_{tmin}(2)_{input} (1)_{chir}\,[10] ,
\eea
where the error budget accounts for the same sources of uncertainties already considered for the charm sector in Sec.~\ref{sec:FMD}. 

Mass ratios can be combined with the experimental values of $B_{(s)}$-meson masses \cite{PDG} to obtain
 \be
      m_{B^*} = 5320.5 ~ ( 7.6) ~ \mbox{MeV} \hspace{5mm} \mathrm{and} \hspace{5mm} m_{B_s^*} = 5411.8 ~ (6.2) ~ \mbox{MeV} ~ ,
 \ee
that compare nicely with the experimental values $m_{B^*}^{exp} = 5324.83 (32)$ MeV and $m_{B_s^*}^{exp} = 5415.4 (1.6)$ MeV \cite{PDG}. 

As for the decay constant ratios, we can compare our results with a recent computation \cite{DavisB} obtained from $N_f = 2 + 1 + 1$ simulations (like the ones considered in this work), $f_{B^*}/f_B = 0.941 (26)$ and $f_{B_s^*}/f_{B_s} = 0.953 (23)$, as well as with a recent determination based on the QCD sum rule approach \cite{Lucha:2015xua}, $f_{B^*}/f_B = 0.944 (23)$ and $f_{B_s^*}/f_{B_s} = 0.947 (30)$. 
All these estimates are nicely consistent with our results. 
On the contrary we find again a $\simeq 10\%$ difference with the $N_f = 2$ determination $f_{B^*}/f_B = 1.051(17)$ from Ref.~\cite{sanfilippo_noi}. 

Eventually, combining our results for the ratios with the pseudoscalar decay constants calculated by ETMC in Ref.~\cite{Bussone:2016iua} yields
 \be
      f_{B^*} = 186.4 ~ (7.1) ~ \mbox{MeV} \hspace{5mm} \mathrm{and} \hspace{5mm} f_{B_s^*} = 223.1 ~ (5.6) ~ \mbox{MeV} ~ .
 \ee

\section{Conclusions}
\label{sec:concl}

We have computed the masses and the decay constants of vector heavy-light mesons using ETMC gauge configurations with $N_f = 2 + 1 +1$ dynamical quarks. 
Our results reproduce very well the experimental values of both $D_{(s)}^*$- and $B_{(s)}^*$-meson masses. 

We have found that $f_{D_{(s)}^*}/f_{D_{(s)}} > 1$ and $f_{B_{(s)}^*}/f_{B_{(s)}} < 1$ with a spin-flavor symmetry breaking effect of $\simeq +8 \%$ in the charm sector and $\simeq - 4 \%$ in the beauty sector. 
Our results for the decay constant ratio exhibit a tension with the corresponding lattice determinations obtained by ETMC at $N_f = 2$ \cite{incriminato,sanfilippo_noi}, while they are consistent with the findings of Refs.~\cite{DavisD,DavisB} obtained by HPQCD with $N_f = 2 + 1 (+1)$ dynamical quarks. 

Since our present analysis follow almost the same steps of the previous ETMC analyses at $N_f = 2$, the observed $\simeq 10\%$ tension may be due to a dependence on the number of sea quarks, and in particular to the inclusion of the strange quark. 
The possibility that the observed difference can be attributed to a quenching effect of the strange quark is a quite interesting issue, because its size would be larger than what typically expected. 
Further investigations at different $N_f$ values are therefore required.

\section*{Acknowledgments}
\footnotesize{We gratefully acknowledge the CPU time provided by PRACE under the project PRA067 on the BG/Q system Juqueen at JSC (Germany) and by CINECA under the specific initiative INFN-LQCD123 on the BG/Q system Fermi at CINECA (Italy).}

\end{document}